\documentclass{nature}
\usepackage{graphicx} 
\usepackage{color}
\usepackage{cite}
\usepackage{xcolor}
\usepackage{caption}

\title{Observation of Unidirectional Bound States in the Continuum
}

\author{Xuefan Yin$^{1,2,3}$, Jicheng Jin$^{1}$, Marin Solja\v{c}i\'{c}$^2$, Chao Peng$^{1,2*}$, \& Bo Zhen$^3$}
\begin{document}
\maketitle

\begin{affiliations}
\item State Key Laboratory of Advanced Optical Communication Systems and Networks, Department of Electronics, Peking University, Beijing 100871, China
\item Department of Physics, Massachusetts Institute of Technology, Cambridge, MA 02139, USA.
\item Department of Physics and Astronomy, University of Pennsylvania, Philadelphia, PA 19104, USA
\end{affiliations}

\begin{abstract}
Unidirectional
radiation is important for a variety of optoelectronic applications. Many unidirectional emitters exist, but they all rely on the use of materials or structures that forbid outgoing waves, i.e. mirrors. Here, we theoretically propose and experimentally demonstrate a class of resonances in photonic crystal slabs, which only radiate towards a single side with no mirror placed on the other side --- we call them ``unidirectional bound states in the continuum". These resonances are found to emerge when a pair of half-integer topological charges in the polarization field
bounce into each other in the momentum space.
We experimentally demonstrate such resonances in the telecommunication regime, where we achieve single-sided quality factor as high as $1.6\times10^5$, equivalent to a radiation asymmetry ratio of $27.7$ dB.
Our work represents a vivid example of applying topological principles to improve optoelectronic devices. Possible applications of our work include grating couplers, photonic-crystal surface-emitting lasers, and antennas for light detection and ranging.
\end{abstract}

Topological defects\cite{mermin_topological_1979}, characterized by quantized invariants, offer a general 
picture to understand many exotic phenomena in the real space such as quantum vortices in superfluids and singular optical beams\cite{gbur_singular_2016}.
It has been recently found that topological defects can also emerge in the momentum space, giving rise to 
interesting consequences.
One such example are the bound states in the continuum\cite{von_neuman_uber_1929} (BICs) in photonic crystal (PhC) slabs: these special resonances reside inside the continuous spectrum of extended radiating modes, yet, counter-intuitively, remain spatially confined and maintain infinitely long lifetimes.

Since initially proposed\cite{von_neuman_uber_1929}, BICs have been found in a variety of wave systems\cite{friedrich_interfering_1985,fan_analysis_2002,Longhi_09,plotnik_experimental_2011,hsu_observation_2013,Longhi_2013,kodigala_lasing_2017,gomis-bresco_anisotropy-induced_2017,molina_surface_2012,carletti_giant_2018,monticone2014embedded,bulgakov_bound_2017,lim_character_1969,cobelli_experimental_2009} and have been widely used in various applications\cite{hirose2014watt,Fan_filter,Chow_sensor}.
Although seemingly unconventional, BICs in PhC slabs are fundamentally vortices in the momentum space, characterized by quantized integer topological charges\cite{zhen_topological_2014,bulgakov_topological_2017}.
The lack of a continuous definition of polarization at the vortex center forbids far-field radiation from this resonance, in the presence of radiation channels, resulting in a BIC.
So far, most BICs realized in PhC slabs rely on spatial symmetries to reduce the number of independent radiation channels.
For example, many BICs are found in up-down mirror symmetric systems\cite{zhen_topological_2014,hsu_observation_2013,ni_tunable_2016,yang_analytical_2014}, where no upward radiation necessarily implies no downward radiation.
On the other hand, whether or not unidirectional BICs --- resonances that only radiate towards a single side with no mirror placed on the other side\cite{Zhou_2016_single_side} --- can exist has remained an open question.

Here we theoretically propose and experimentally demonstrate unidirectional BICs enabled by topological charges in the polarization field of PhC slab resonances.
Specifically, we first split the integer topological charge carried by a BIC, by breaking certain spatial symmetries, into a pair of half-integer topological charges: they each correspond to a circularly-polarized resonance in the downward far-field radiation.
As the structure is continuously varied, the two half-integer topological charges keep evolving in the momentum space until they bounce into each other and, again, act like an integer charge.
At this point, downward radiation from this resonance is disallowed as its far-field polarization is undefined --- this is the topological nature of unidirectional BICs.
We further fabricate PhC samples and experimentally demonstrate unidirectional BICs by directly proving the vanishing of their downward far-field radiation.

We start by showing that unidirectional BICs can be achieved by first splitting and then merging a pair of half-integer topological charges of polarization long axes in the momentum space.
As a specific example, we consider a 1D-periodic PhC slab where infinitely long bars with gaps of $w=358$ nm are defined in a $500$ nm-thick silicon layer with refractive index of $n=3.48$ at a periodicity of $a = 772$ nm (Fig.~1a-d).
Both the top and bottom silica cladding layers ($n=1.46$) are assumed to be semi-infinitely thick.
When the sidewalls of the bars are vertical ($\theta = 90^{\circ}$, Fig.~1b), the PhC slab is up-down and left-right symmetric,
and a BIC is found on a TE-like band (TE1) along the $k_x$ axis off the normal direction at $k_{x}a/2\pi = 0.176$.
In this up-down symmetric structure, the radiative decay rate of a mode towards the top ($\gamma_{\rm t}$, orange) is always the same as that towards the bottom ($\gamma_{\rm b}$, blue), both of which reduce to 0 at the BIC (middle panels, Fig.~1b).
Fundamentally, this BIC can be understood as a topological defect in the far-field polarization long axes 
that carries an integer topological charge of $q=+1$, as the polarization long axis winds around the BIC by $2\pi$ in the counter-clockwise (CCW) direction  (yellow arrows, bottom panel, Fig.~1b).

As one of the sidewalls is tilted away from being vertical ($\theta = 81^{\circ}$, Fig.~1c), the PhC slab is no longer up-down symmetric, so $\gamma_{\rm t}$ and $\gamma_{\rm b}$ are no longer simply related. Importantly, no BIC exists in this structure any more: the radiative decay rate towards either the top or bottom, $\gamma_{\rm t,b}$, never reaches 0 (middle panels, Fig.~1c).
The integer charge of the BIC has now split into a pair of half-integer charges $q=+1/2$, each being a circularly-polarized resonance (bottom panel, Fig.~1c).
The two half-integer charges are related to each other by the $y$-mirror symmetry of the structure, which also guarantees these two circularly polarized resonances to be opposite in helicity: CCW for one (red) and clockwise (CW) for the other (green).

As the sidewalls are further tilted, 
the two circularly-polarized resonances in downward radiation keep moving in the momentum space following trajectories shown in Fig.~1e: red for CCW and green for CW.
There are no unidirectional BICs in the system until $\theta$ is decreased to $75^\circ$ (Fig.~1d) and the CW and CCW trajectories meet on the $k_x$ axis.
At this point, any downward radiation needs to be both CW- and CCW-circularly polarized at the same time, which can never be satisfied as the two polarization states are orthogonal to each other.
As a result, this mode cannot have any downward radiation even though no mirror is placed on the bottom and the radiation channel is open to it.
This interpretation agrees with numerical simulation results, where $\gamma_{\rm b}$ reaches 0 while $\gamma_{\rm t}$ remains finite --- we name such resonances ``unidirectional BICs" (middle panels, Fig.~1d).
From the viewpoint of topology, unidirectional BICs can be understood as the merging point between two half-integer charges, where they act like an integer charge, forbidding any radiative loss.
See Supplementary Information Section \uppercase\expandafter{\romannumeral1}- \uppercase\expandafter{\romannumeral3}  for more details.

Following our interpretation, unidirectional BICs can be designed by bouncing two half-integer topological charges into each other while each representing a circularly-polarized resonance with opposite helicities.
This mechanism does not depend on any structural symmetry and can occur when the top and bottom cladding materials are different.
Next, we present our unidirectional BIC design with air on the top and SiO$_{2}$ on the bottom, which we fabricate and experimentally characterize later.
The PhC slab consists of a periodic array of one-dimensional bars defined in a $500$ nm-thick silicon-on-insulator (SOI) wafer at the periodicity of $a = 825$ nm (left panel, Fig.~2a).
The sidewalls are tilted to specific angles, $\theta_{\rm L} = 79 ^\circ$ and $\theta_{\rm R} = 75 ^\circ$, to achieve a unidirectional BIC: as shown its mode profile of $E_y$ component, downward radiation $\gamma_{\rm b}$ is significantly lower, by over 70 dB, than its upward radiation $\gamma_{\rm t}$ (right panel, Fig.~2a).
The asymmetry ratio between up- and downward radiation intensity, $\eta = \gamma_{\rm t}/\gamma_{\rm b}$, is calculated for different $k$-points (color map, Fig.~2b), where the extremely bright spot marks the location of the unidirectional BIC at $k_x a /2\pi = 0.0854$.
A line-cut of the color map along the $k_x$ axis shows the asymmetry ratio $\eta$ diverging into infinity, which is the characteristic feature of unidirectional BICs (Fig.~2c).
Overlaid on the color map in Fig.~2b is the polarization long axes plot for the downward far-field radiation from nearby resonances; indeed, an integer winding of the polarization long axes, $q=+1$, is observed around the unidirectional BIC, which is consistent with our interpretation presented in Fig. 1d.

To verify our theoretical findings, we fabricate PhC samples with unidirectional BICs using plasma-enhanced chemical vapor deposition, e-beam lithography, and reactive ion etching (RIE) processes.
The scanning electron microscope images are shown in Fig.~3a,b.
Briefly, a thermal SiO$_2$ layer with thickness of approximately $110$ nm is first deposited on the wafer as the hard mask.
Different from standard RIE processes that use horizontal substrates, our sample is placed on a wedged substrate that allows us to etch the silicon layer at a slanted angle; as a result, high-quality air gaps with tilted sidewalls are achieved (Fig.~3b).
Because of the shadowing effect, the angles of the left and right sidewalls are not identical: $\theta_{\rm L} = 79 ^\circ$ and $\theta_{\rm R} = 75 ^\circ$.
The width of the air gaps $w$ is swept from $340$ to $360$ nm, at the step of $2.5$ nm
to best capture the unidirectional BIC design at $w=$352 nm.
See Methods and Supplementary Information Section \uppercase\expandafter{\romannumeral4} for more details about the fabrication.

To demonstrate the existence of unidirectional BICs, up- and downward radiative decay rates from our fabricated samples are independently characterized using the experimental setup\cite{Isarael2005PhysRevB } schematically shown in Fig.~3c.
A tunable telecommunication laser in the C+L-band is first sent through a polarizer in the $x$-direction (POL) before it is focused by a lens (L1) onto the rear focal plane (RFP) of an infinity-corrected objective lens.
To achieve on-resonance coupling condition, for each excitation wavelength $\lambda$, the incident angle is tuned, by moving L1 in the $x-y$ plane, to excite a resonance of the sample.
Each excited resonance radiates towards the top and bottom according to its radiative decay rate into each channel respectively.
Upward (downward) far-field radiation from this resonance is then collected by the con-focal setup shown on the right (left), marked with a orange (blue) dashed box, where the beam is shrunk by 0.67 times through a $4f$ system to best fit the camera.
Our on-resonance excitation scheme here is similar to previously reported results\cite{regan_direct_2016,zhou_observation_2018,galli_light_2009}. See Methods and Supplementary Information Section \uppercase\expandafter{\romannumeral5} for more details.

As an example, the experimental comparison between up- and downward radiation from a resonance at $\lambda = 1551$ nm is presented in Fig.~4a.
Here, the excitation laser is on resonance with a mode on the $k_x$ axis at $k_y a/2\pi=0.01$.
Labeling of the momentum spaces is calibrated with respect to the known numerical aperture of the objectives (NA=$0.26$), shown as white circles.
The characteristic feature of the unidirectional BIC --- marked by a white arrow on the $k_x$ axis  --- is qualitatively shown in the comparison between the two figures: for resonances near the unidirectional BIC, downward radiation (X$^{'}$, Y$^{'}$, Z$^{'}$) is always much weaker than upward radiation (X, Y, Z).
On the other hand, for resonances on the left half of the momentum space that are far away from the unidirectional BIC, their up- and downward radiations are comparable.

A more quantitative demonstration of the unidirectional BIC is achieved by measuring the asymmetry ratio of the resonances. Two movable pinholes (not shown in Fig.~3c) with diameters of $300$ $\mu$m are placed at the image planes of the objectives' RFPs to select specific $k$ points. Three examples are shown in Fig.~4b, where upward (X, Y, Z) and downward (X$^{'}$, Y$^{'}$, Z$^{'}$) radiation intensities are measured as the excitation wavelength scans through the three resonances.
As expected, all reflection and transmission spectra exhibit symmetric Lorentzian features\cite{regan_direct_2016}:
the excitation efficiency reaches its maximum when the excitation is on resonance, which happens at $\lambda = 1553.7$, $1551.2$, and $1549.4$ nm respectively.
Accordingly, both central wavelengths $\lambda$ and quality factors $Q_{\rm tot}$ of the resonances can be extracted by fitting the experimental results to Lorentzian functions.
Repeating this procedure for all resonances along the X-Z line, we achieve a good agreement between experimentally-extracted resonance wavelengths (red crosses) and numerical simulation results (blue line, Fig.~4c).

To demonstrate the existence of unidirectional BICs, we further experimentally extract the downward radiative decay rate of the resonances $\gamma_{\rm b} = \omega/Q_{\rm b}$ and show it reduces to 0 at certain $k$ points.
Here, $\omega$ is the resonance frequency and $Q_{\rm b}$ is the radiative quality factor that only accounts for the downward radiation.
In practice, the observed total loss $\omega/Q_{\rm tot}$ is composed of 
non-radiative loss due to absorption, scattering, lateral leakage\cite{liang_three-dimensional_2012,ni_analytical_2017}, and radiative losses towards the top and bottom. As these resonances are close in the momentum space and share similar mode profiles, it is reasonable to assume they share a similar non-radiative decay rate, which is found to be $Q_{\rm non-rad} = 2080$ through numerical fitting (see Supplementary Information Section \uppercase\expandafter{\romannumeral6} for  details).
Up- and downward radiative decay rates can be separated based on measured asymmetry ratio $\eta$. Our experimentally extracted $Q_{\rm b}$ are presented in Fig.~4c as red crosses, which show a good agreement with the numerical simulation results (blue line).
In particular, the fact that the bottom radiation $\gamma_{\rm b}$ reduces to almost 0 as $Q_{\rm b}$ diverges at $k_{x}a/2\pi = 0.088$ proves our demonstration of the unidirectional BIC.

At the unidirectional BIC (Y, Y$^{'}$), the experimentally measured asymmetry ratio between up- and downward radiation reaches a maximum of $27.7$ dB, meaning over $99.8\%$ of light radiates towards the top (Fig.~5a).
Such highly directional radiation is essential and desirable in many optoelectronic devices.
For example, grating couplers\cite{Roncone93_grating_coupler,Taillaert04_grating_coupler,Vermeulen_10_grating_coupler,markwade_grating_coupler,Jelena_grating_coupler}, which couple light between the nanophotonic waveguides and fibers, are among the most important elements in photonic integrated circuits. While having been studied extensively, the performance of grating couplers is still far from optimal: one of the major challenges is the insertion loss that is now mainly limited by unwanted downward radiation losses\cite{Jelena_grating_coupler}.
Accordingly, the unidirectional BICs we demonstrate here provide a practical and effective method to solve this problem by naturally eliminating the downward radiative loss.
Near unidirectional BICs, strong suppression of downward radiation is achieved across a broad bandwidth: over 90\% of the radiation energy is maintained towards the top within a $26$ nm bandwidth from 1536 to 1562 nm (Fig.~5a).
Furthermore, we achieve robust suppression of downward radiation at different out-coupling angles between $5^{\circ}$ and $11^{\circ}$ (Fig.~5b).
Finally, our design is found to have good tolerance to fabrication errors.
See Supplementary Information Section \uppercase\expandafter{\romannumeral6} and \uppercase\expandafter{\romannumeral7} for details.

To summarize, we  propose and  demonstrate a type of resonances, named ``unidirectional BICs", which only radiate towards a single side even though no mirror is placed on the other side.
From the viewpoint of topology, such resonances are induced by first splitting and then merging two half-integer topological charges of polarization long axes.
We fabricate PhC samples on SOI wafers using a modified RIE process, \textcolor{black}{ and measure the upward and downward radiation intensity directly}.
By showing the downward radiation vanishes, we demonstrate the existence of unidirectional BICs,
which are potentially useful for many applications including grating couplers, photonic-crystal surface-emitting lasers, and antennas for light detection and ranging.

\bibliographystyle{naturemag}
\bibliography{reference}


\clearpage

\captionsetup[figure]{labelformat=empty}
\clearpage
\begin{figure}
\centering
\includegraphics[width=8.9cm]{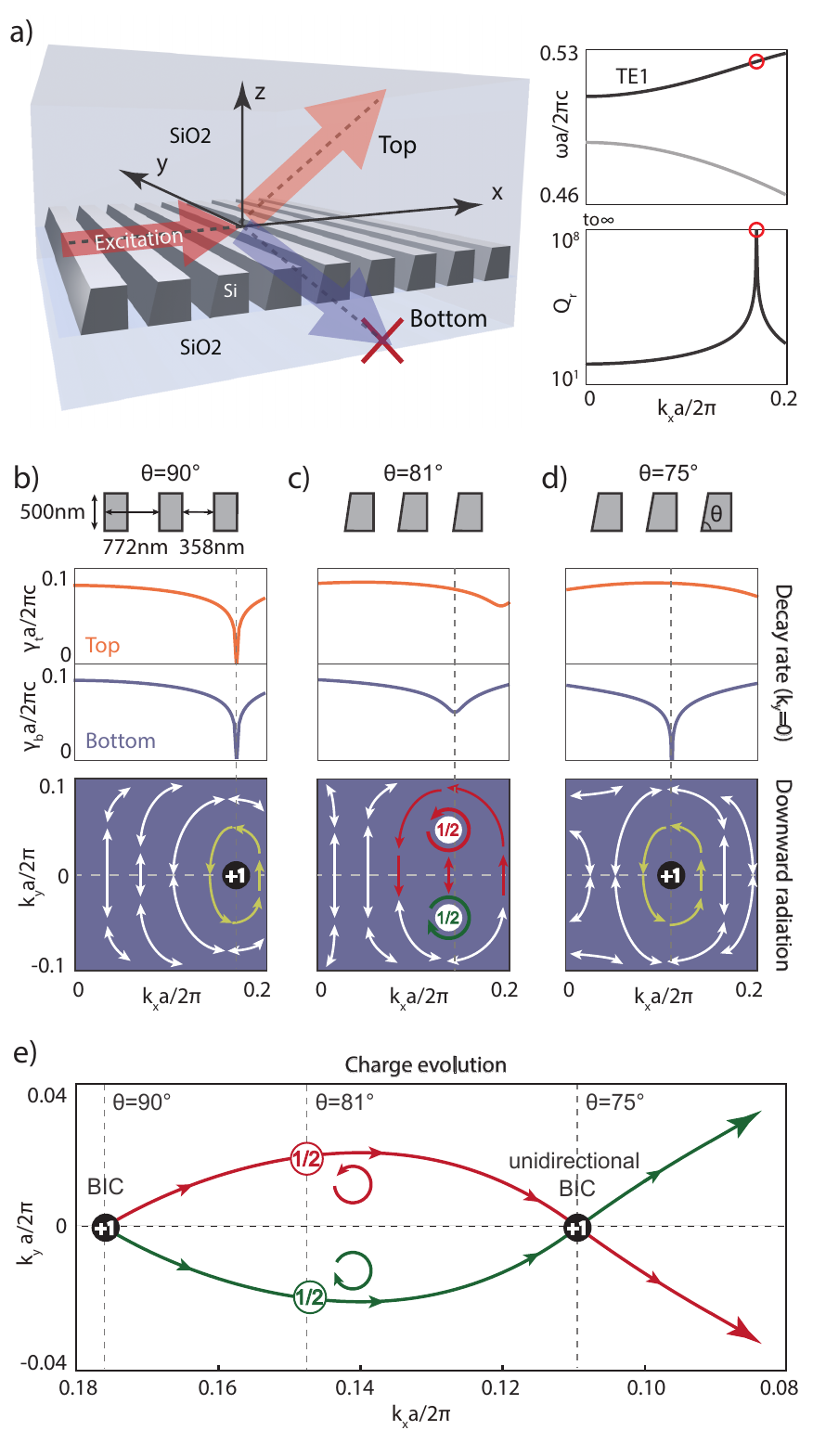}
\caption{\bf{Fig. 1$\mid$ Topological nature of unidirectional BICs}.}
\end{figure}
\clearpage
\begin{figure}
\centering
\captionsetup{labelformat=empty}
\caption{{\bf a,} Schematic of a PhC slab.
When the PhC slab is up-down symmetric, a bound state in the continuum (BIC) is found on the TE$_{1}$ band, marked with a red circle, at which point the quality factor $Q_r$ diverges to infinity.
{\bf b,}
Due to up-down symmetry, radiative losses from resonances towards the top ($\gamma_{\rm t}$, orange line) are always equal to that towards the bottom ($\gamma_{\rm b}$, blue line), both of which reduce to 0 at the BIC.
The polarization long axis winds around the BIC in the $k-$space, featuring a topological charge of $+1$.
{\bf c,}
The up-down mirror symmetry is broken as the sidewall angle $\theta$ is tuned from $90^\circ$ to $81^\circ$, where BIC no longer exists in the system. Instead the $+1$ charge in the downward radiation polarization splits into a pair of half-integer charges of $+1/2$, marked with red and green circles.
{\bf d,}
As $\theta$ is further tuned to $75^\circ$, the two $+1/2$ charges bounce into each other on the $k_{x}$ axis. At this point, the radiation loss towards the bottom is eliminated ($\gamma_{\rm b} = 0$, blue line), while radiation towards the top ($\gamma_{\rm t}$, orange line) remains finite --- we call this resonance a ``unidirectional BIC".
{\bf e,} The trajectories traced by the two $+1/2$ topological charges (red and green) in the $k-$space as the sidewall angle varies from $90^{\circ}$ to $75^{\circ}$ and beyond.
}
\end{figure}

\clearpage
\begin{figure}
\centering
\includegraphics[width=8.9cm]{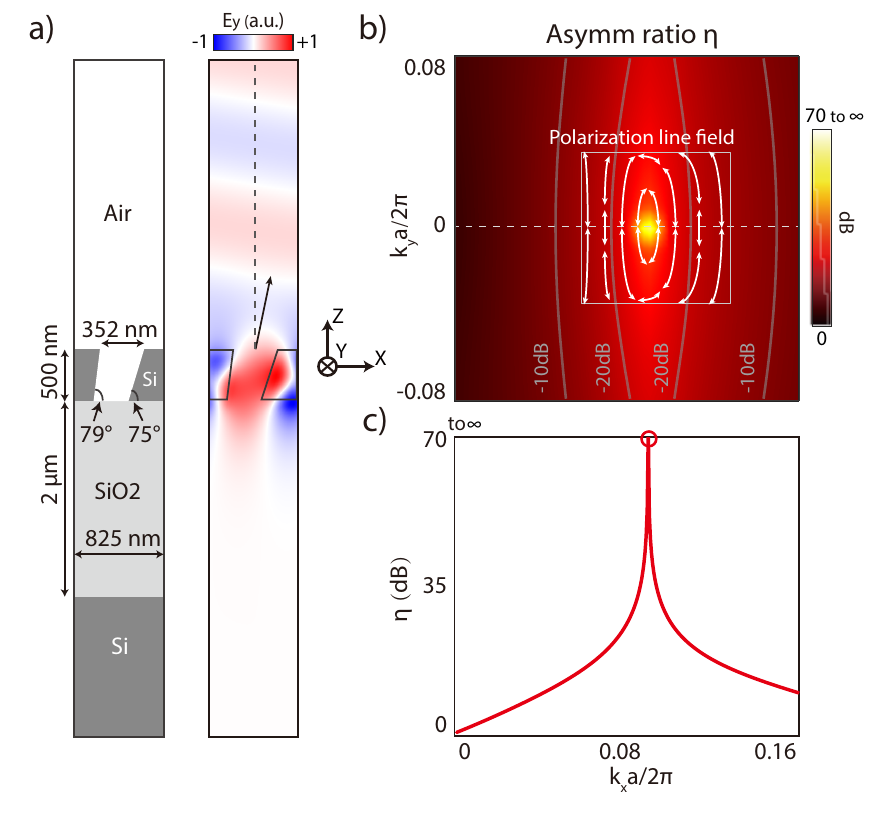}
\caption{{\bf Fig. 2 $\mid$ Numerical confirmation of a unidirectional BIC and sample design.} {\bf a,} Unit cell design of a PhC slab, which supports a unidirectional BIC with its mode profile ($E_{y}$) shown on the right: no downward radiation is observed while upward radiation remains finite.
{\bf b,} The asymmetry ratio between up- and downward radiation loss, $\eta = \gamma_{\rm t}/\gamma_{\rm b}$, diverges to infinity at the unidirectional BIC and remains high in the $k-$space nearby.
Overlaid on the color map are the polarization long axes of the resonances, showing a topological charge of $q=+1$ at the unidirectional BIC --- consistent with Fig.~1d.
{\bf c,} Asymmetry ratio $\eta$ along the $k_{x}$ axis, which diverges into infinity at the unidirectional BIC (marked with a red circle).}
\end{figure}

\clearpage
\begin{figure}
\centering
\includegraphics[width=18.3cm]{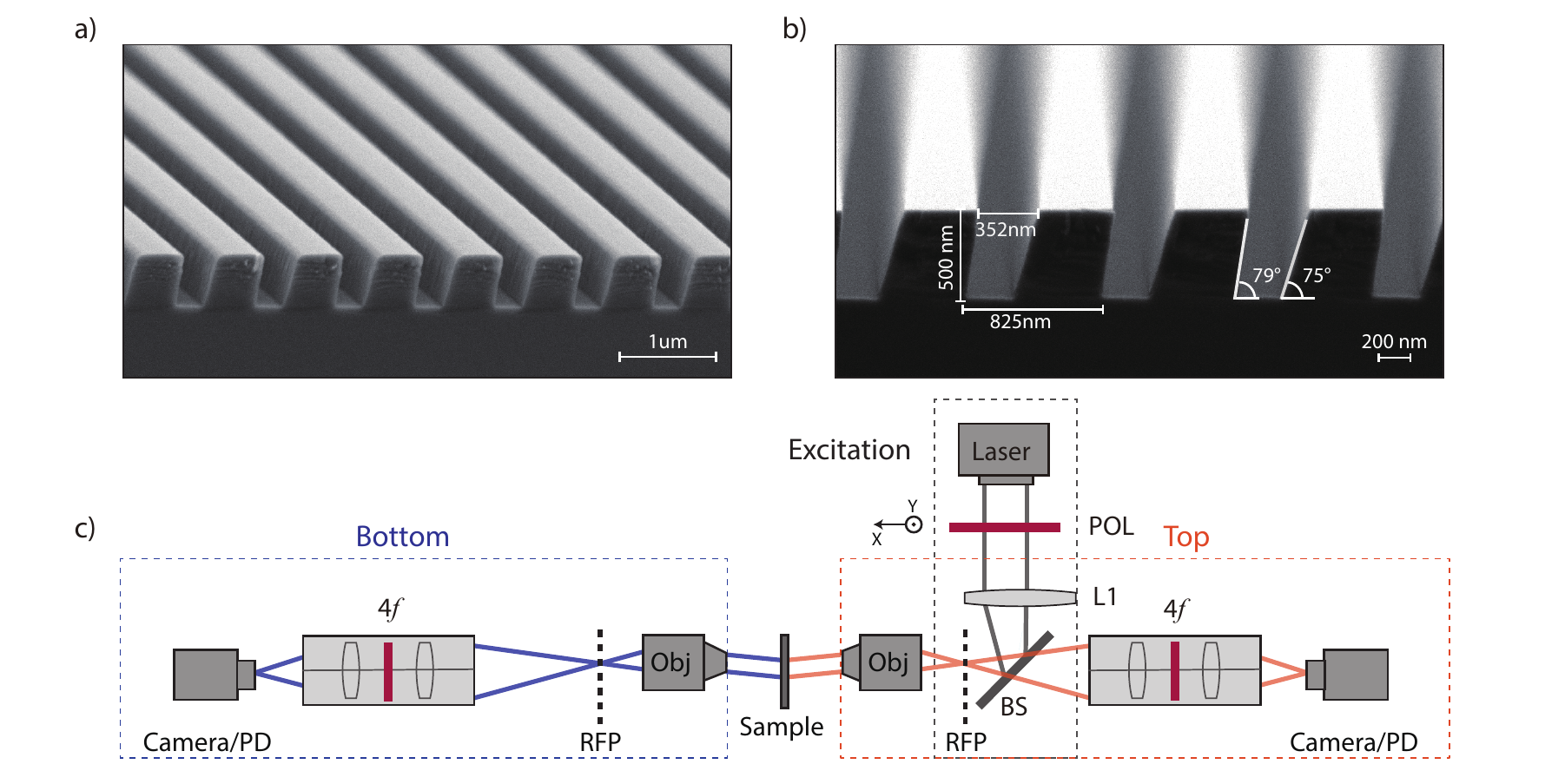}
\caption{{\bf Fig. 3 $\mid$ Fabricated sample and experimental setup.}
{\bf a,b} Scanning electron microscope images of the fabricated PhC sample, corresponding to Fig.~2a, from tilted and side views.
{\bf c,}
Schematic of the setup to independently measure up- and downward radiation intensity from resonances in the PhC sample.
L, lens; Obj, objective; RFP, real focal plane; PD, photodiode; POL, polarizer; BS, beamsplitter; $4f$, relay 4-$f$ optical system.}
\end{figure}

\clearpage
\begin{figure}
\centering
\includegraphics[width=18.3cm]{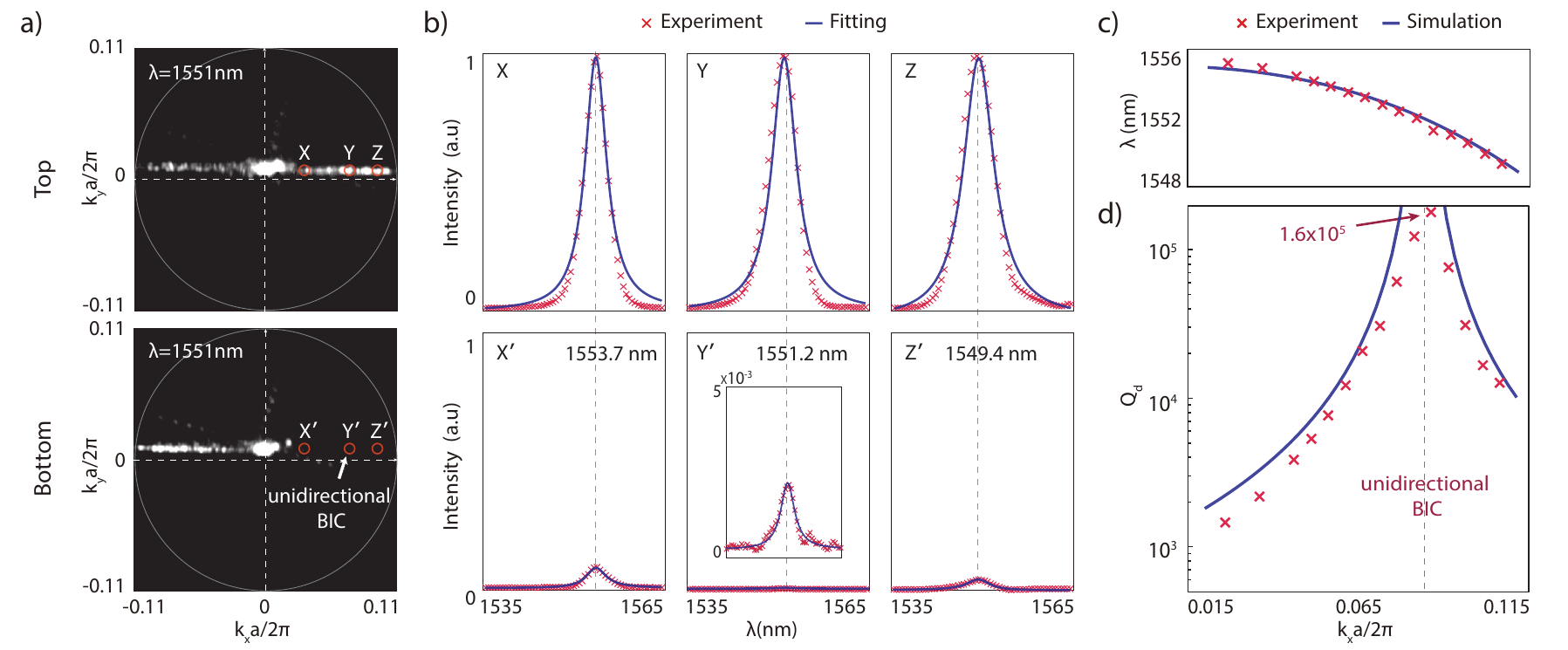}
\caption{{\bf Fig. 4 $\mid$ Experimental observation of unidirectional BICs.}
{\bf a,}
Up- and downward radiation intensities from resonances under the excitation wavelength of $1551$ nm.
In the vicinity of unidirectional BIC on the $k_x$ axis, marked by a white arrow, downward radiation intensities (X$^{'}$, Y$^{'}$, Z$^{'}$) are significantly suppressed compared to the upward (X, Y, Z).
{\bf b,}
Up- and downward radiation intensities from three example resonances as the excitation wavelength scans from $1535$ to $1565$ nm.
{\bf c,d,}
Experimentally extracted band structure and $Q_{\rm b}$ (red crosses), the radiative quality factor that only accounts for downward radiation, showing good agreements with numerical simulation results (blue line).
}
\end{figure}

\clearpage
\begin{figure}
\centering
\includegraphics[width=10 cm]{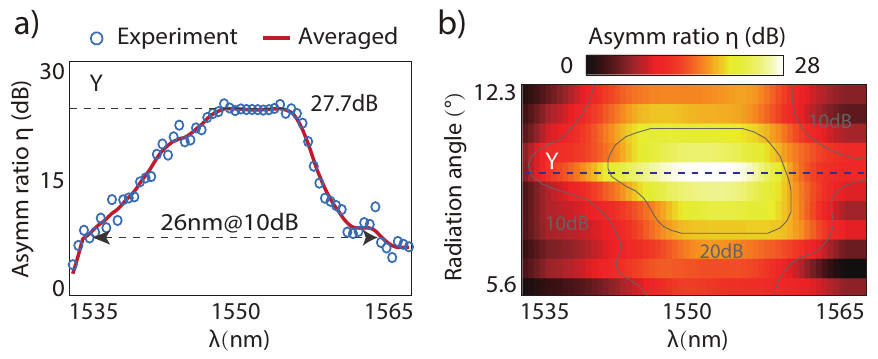}
\caption{{\bf Fig. 5 $\mid$ Projected performance of unidirectional BICs as grating couplers.}
{\bf a,}
Asymmetry ratio $\eta$ between up- and downward radiation intensities for a fixed out-coupling angle of $9^\circ$. The maximum reaches $27.7$ dB near the unidirectional BIC and remains above $10$ dB over a bandwidth of $26$ nm.
{\bf b,}
Highly directional emission is observed  over a wide range of excitation wavelengths and for different out-coupling angles}
\end{figure}
\end{document}